# Radiation-Hard Optical Link for SLHC

K.K. Gan, W. Fernando, H. Kagan, R. Kass, A. Law, S. Smith

Department of Physics, The Ohio State University, Columbus, OH 43210, USA

M.R.M. Lebbi, P.L. Skubic

Department of Physics and Astronomy, University of Oklahoma, Norman, OK 73019, USA

**Abstract**. We study the feasibility of fabricating an optical link for the SLHC ATLAS silicon tracker based on the current pixel optical link architecture. The electrical signals between the current pixel modules and the optical modules are transmitted via micro-twisted cables. The optical signals between the optical modules and the data acquisition system are transmitted via radiation-hard/low-bandwidth SIMM fibres fusion spliced to radiation-tolerant/medium-bandwidth GRIN fibres. The link has several nice features. We have measured the bandwidths of the micro twisted-pair cables to be ~ 1 Gb/s and the fusion spliced fibre ribbon to be ~ 2 Gb/s. We have irradiated PIN and VCSEL arrays with 24 GeV protons and find the arrays can operate up to the SLHC dosage. We have also demonstrated the feasibility of fabricating a novel opto-pack for housing VCSEL and PIN arrays with BeO as the substrate.

## I. INTRODUCTION

The Super LHC (SLHC) is designed to increase the luminosity of the Large Hadron Collider (LHC) by a factor of ten to $10^{35}$ $cm^{-2}s^{-1}$. Accordingly, the radiation level at the detector is expected to increase by a similar factor. The increased data rate and radiation level will pose new challenges for a tracker situated close to the interaction region. The present optical link [1] of the ATLAS pixel detector [2] is mounted on a patch panel instead of directly on a pixel module. This separation greatly reduces the radiation level at the optical modules (opto-boards) and simplifies the design and production of both the pixel modules and opto-boards. Data communication between the separated modules is achieved by transmitting electrical signals using ~ 1 m of micro twisted-pair cables. The optical signals between each opto-board and the off-detector optical electronics are then transmitted via 8 m of radiation-hard/low-bandwidth SIMM (Step Index Multi-Mode) fibre ribbon fusion spliced to 70 m of radiation-tolerant/medium-bandwidth GRIN (Graded Index) fibre ribbon. The optical signals are generated and received using the VCSEL (Vertical Cavity Surface Emitting Laser) and PIN arrays, respectively. We currently transmit optical signals to the detector at 40 Mb/s and from the detector at 80 Mb/s by using both edge of the 40 MHz clock. For the SLHC, the required bandwidths are ~ 0.160 and 1 Gb/s. If the present architecture can transmit signals at the higher speed, the constraint of requiring no extra service space is automatically satisfied.

We have started an R&D program to study the feasibility of an upgrade based on the optical link architecture of the current pixel detector while taking advantage of the several years of R&D effort and production experience. In this paper, we present results on the bandwidth measurement of micro twisted-pair cables and fusion spliced SIMM/GRIN fibres and the radiation hardness of VCSEL and PIN arrays. In addition, we report the results on a novel opto-pack for housing VCSEL and PIN arrays fabricated with BeO as the substrate.

## II. BANDWIDTH OF MICRO TWISTED-PAIR CABLES

Commercial copper cables [3] can transmit several Gb/s over tens of meters. However, the diameters of these cables are too large for the pixel detector. The present pixel optical link uses a micro twisted-pair of wires for transmission of low voltage differential signals (LVDS) between a pixel module and the driver and receiver chips on an opto-board. Each pair of wires is twisted 5 turns per inch (TPI) which corresponds to 2 turns per cm. For barrel pixel detectors, each wire is aluminium with a diameter of 100 µm (38 AWG) plus 25 µm of insulation, for an outer diameter of 150 µm. The length of the twisted pairs varies from 81 to 142 cm. The wires for the endcap pixel detector are finer, 60 µm with 12 µm of insulation. The length of these copper twisted pairs is ~ 80 cm. The impedance of the twisted pairs is ~ 75 Ω.

We have measured the bandwidths of micro twisted-pairs of various lengths, diameters, and numbers of turns per cm [4]. We transmitted LVDS pseudo-random data in the selected cable and measured the signal characteristics at the termination with a LeCroy WaveMASTER 8600A (6 GHz) oscilloscope and differential probe (7.5 GHz). The rise and fall times of the cables are shown in Figure 1. The thickest cable tested with 25 µm of insulation and 2 turns/cm has the fastest rise time. However, the current barrel cable which is slightly thinner has similar performance.

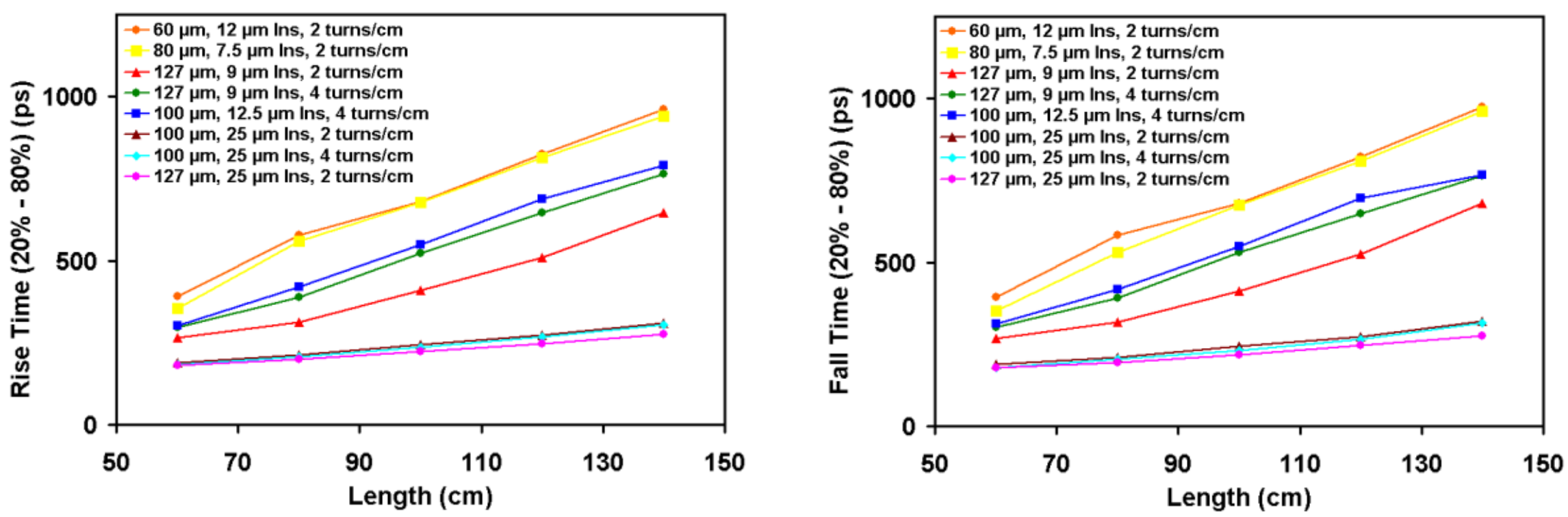

Figure 1: The rise and fall times (20-80%) of the micro twisted-pairs vs. wire length for wires of various diameters and turns per cm.

The eye diagrams produced by transmitting pseudo-random data of 640 Mb/s and 1,280 Mb/s in the current barrel cable and the thicker cable are shown in Figure 2. The masks shown are adapted from Figure 39-5 and Table 39-4 of the Gigabit Ethernet Specification (IEEE Standard 802.3) with the mask voltage levels modified to match the LVDS receiver chip used. From these figures, it is evident that the micro-twisted cables are adequate for transmitting signals at 640 Mb/s and that transmission at 1,280 Mb/s might be acceptable. The thicker cable has a slightly higher bandwidth with the slightly more open eye diagrams.

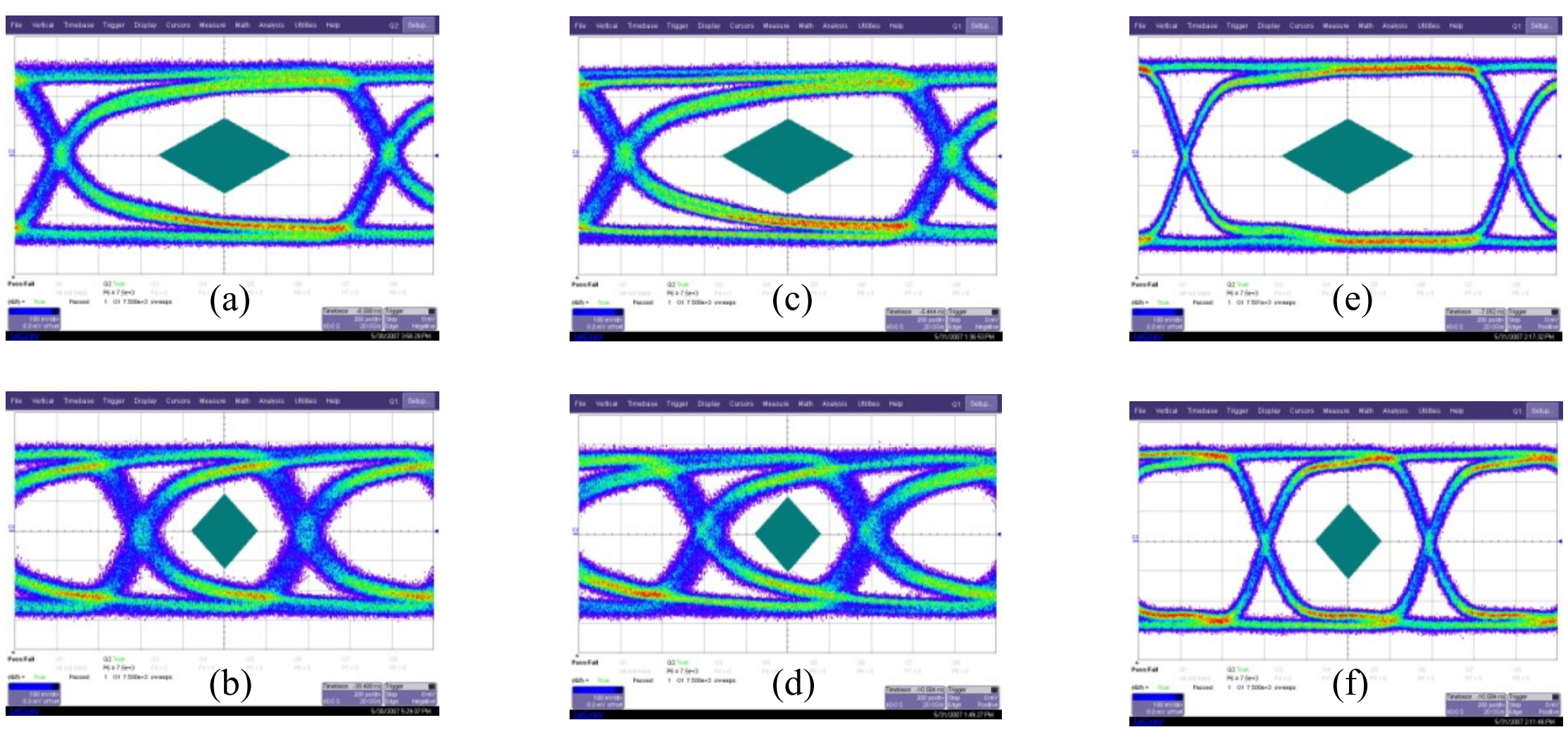

Figure 2: Eye diagrams for signals of (a) 640 and (b) 1,280 Mb/s for a 1.4 m cable with a diameter of 127 µm plus 25 µm of insulation, twisted 2 turns/cm. (c,d) show the corresponding signals for the pixel barrel cable of the same length. (e,f) show the corresponding signals for the pixel barrel cable of 60 cm.

## III. BANDWIDTH OF FUSION SPLICED SIMM/GRIN FIBRE

The optical link of the present pixel detector uses 8 m of SIMM fibre ribbons fusion spliced to 70 m of GRIN ribbon. The former has low bandwidth with a radiation-hard pure silica core. The latter has medium bandwidth and is radiation tolerant. The mix use of fibers was due to the limited supply of the more costly SIMM fiber. The signal is transmitted from the pixel modules to the counting room using 50 μm SIMM fibre ribbons fusion spliced to 62.5 μm GRIN fibre ribbons to reduce the signal lost near the interface. For the transmission in the other direction, 50 μm GRIN fibre ribbons were used.

We have measured the bandwidth of 8 m of 50 μm SIMM fibre ribbon fusion spliced to 80 m of 62.5 μm GRIN fibre. The ribbon contains eight fibers. The optical signal was generated with an 850 nm VCSEL, contained within a Finisar FTRJ-8519-1-2.5 fibre optic transceiver, and measured using the above oscilloscope with a 4.5 GHz optical to electrical converter. Resultant eye diagrams for pseudo-random signals of 1 and 2 Gb/s for a fiber are shown in Figure 3. The mask shown is adapted from Figure 38-2 of the Gigabit Ethernet Specification (IEEE Standard 802.3) and in accordance with the specification, a fourth-order Bessel-Thomson software filter is used to view the signals. We use an 1.5 Gb/s filter for the 2 Gb/s transmission as recommended. For the 1 Gb/s transmission, we choose the same bandwidth instead of the recommended 750 Mb/s to show the rise and fall times of the signal. It is evident from these results that the fibre can adequately transmit signals up to 2 Gb/s and hence the transmission bandwidth of the wire link will be the limiting factor in the present pixel detector transmission lines.

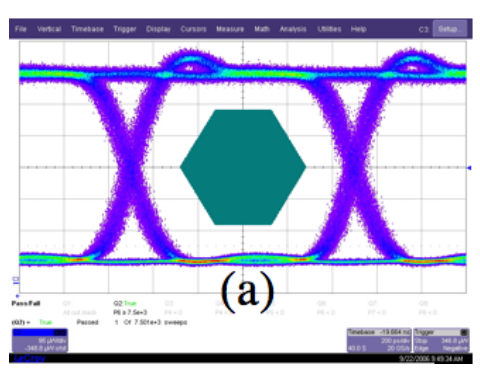

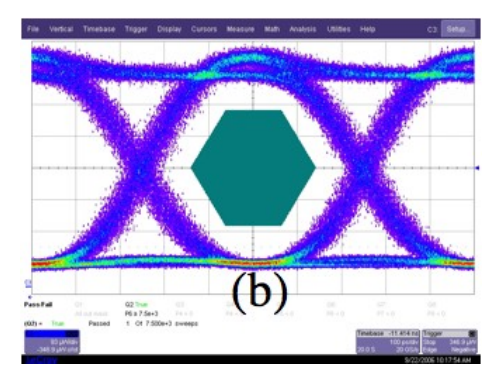

Figure 3: Eye diagrams for optical signals of (a) 1 and (b) 2 Gb/s in a fusion spliced SIMM/GRIN fibre.

## IV. RADIATION HARDNESS OF PIN AND VCSEL ARRAYS

We use the Non Ionizing Energy Loss (NIEL) scaling hypothesis to estimate the SLHC fluences [5-7] at the present pixel optical link location (PP0). The estimate is based on the assumption that the main radiation effect is bulk damage in the VCSEL and PIN with the displacement of atoms. After five years of operation at the SLHC (3,000 $fb^{-1}$), we expect the silicon component (PIN) to be exposed to a maximum total fluence of 1.5 x $10^{15}$ 1-MeV $n_{eq}/cm^2$ [8]. The corresponding fluence for a GaAs component (VCSEL) is 8.2 x $10^{15}$ 1-MeV $n_{eq}/cm^2$. We study the response of the optical link to a high dose of 24 GeV protons. The expected equivalent fluences at SLHC are 2.6 and 1.6 x $10^{15}$ $p/cm^2$ for a silicon and GaAs device, respectively. For simplicity, we present the results with dosage in Mrad, 69 and 34 Mrad.

We irradiated four opto-boards with PIN and VCSEL arrays from various vendors using 24 GeV protons at the T7 facility at CERN [9]. Each board was instrumented with one silicon PIN array and a pair of GaAs VCSEL arrays. The PIN arrays were all fabricated by one vendor, Truelight, and the VCSEL arrays were fabricated by three vendors, Optowell (AM85-1N112), Advanced Optical Components (HFE8012-101), and ULM Photonics (two varieties, 5 and 10 Gb/s, ULM850-05-TN-B0112U and ULM850-10-TN-N0112U). On the opto-boards, each of the PIN and VCSEL arrays coupled to radiation-hard ASICs produced for the current pixel optical link [1], the DORIC (Digital Opto Receiver Integrated Circuit) and VDC (VCSEL Driver Chip). Furthermore, the opto-boards were mounted on a shuttle system which enabled us to easily move in and out of the beam for annealing of the VCSEL arrays.

The test system monitored various parameters of the opto-boards throughout the irradiation. Of particular interest was the optical power of the VCSEL arrays vs. dosage as shown in Figure 4. The power decreased during the irradiation as expected. We annealed the arrays by moving the opto-boards out of the beam and passing the maximum allowable current (up to 20 mA per channel) through the arrays for several hours each day. The optical power increased during the annealing. However, there was insufficient time for a complete annealing. Consequently both the A.O.C. and ULM 5 Gb/s arrays had no optical power at ~ 50

Mrad. The ULM 10 Gb/s arrays continued to have optical power of more than 100 μW up to 65 Mrad and the Optowell arrays survived up to at least 70 Mrad. We believe that more arrays would have survived if we had more time for annealing. However, it should be emphasized that all arrays continue to produce decent optical power up to the SLHC dosage of 34 Mrad.

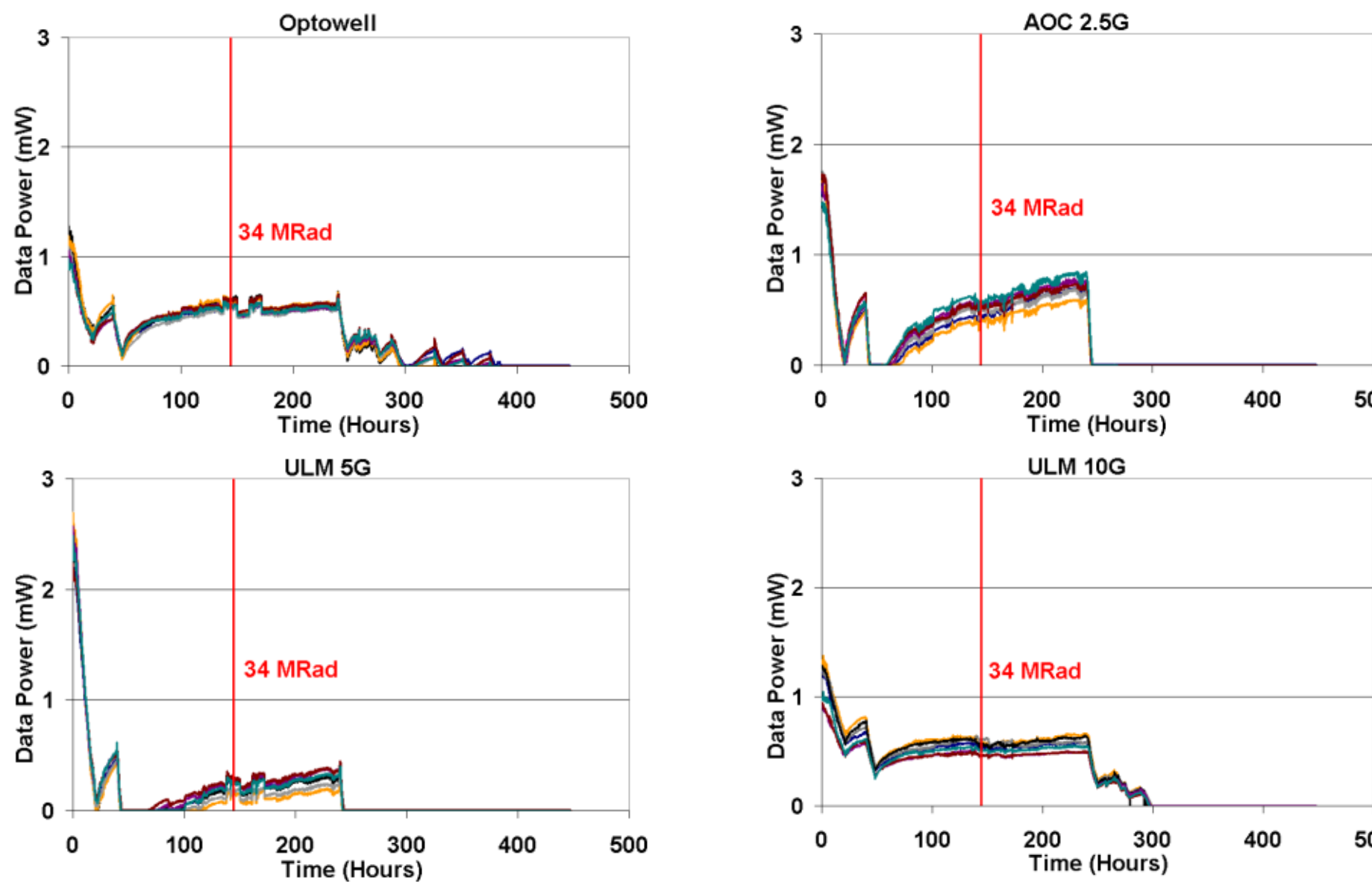

Figure 4: Optical power as a function of time (dosage) of eight VCSEL arrays. The power decreased during the irradiation but increased during the annealing as expected.

The silicon PIN arrays survived the irradiation quite well. After 115 Mrad, the responsivities decreased to ~ 35% of the pre-irradiation level as shown in Figure 5. No degradation in the rise/fall times (20-80%) has been observed (~ 1 ns).

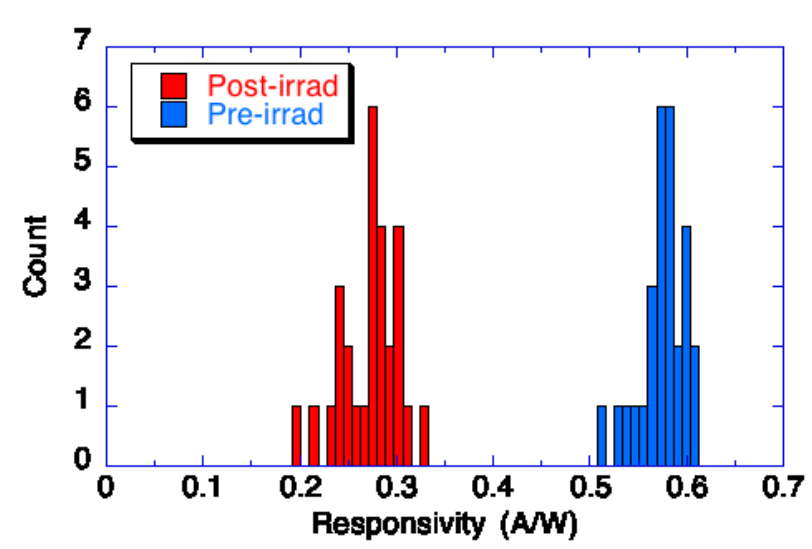

Figure 5: Responsivities of the four PIN arrays before (blue) and after (red) irradiation.

## V. Novel optical package

We have developed a new novel package for the VCSEL and PIN arrays as shown in Figures 6 and 7. The package is as compact as the package provided by Academic Sinica, Taiwan for the current ATLAS pixel optical link [10]. However, the base is fabricated using BeO instead of PCB for much better removal of the heat produced by the VCSEL which is the major heat source in the opto-link. The through hole vias for connecting to the anode and cathode pads on an array are replaced by three dimensional traces that go over the edge of the BeO base. Wire bonds connect the driver (receiver) chip to the VCSEL (PIN) array. This avoids the need for the challenging soldering of the micro-leads (250 μm width) to the BeO opto-board as it is difficult to supply sufficient heat to a tiny lead to attach to a trace on an excellent conductor. Moreover, the traces can be rewire-bonded for diagnostics and rerouting purposes as sometimes needed, especially during the R&D phase of a project.

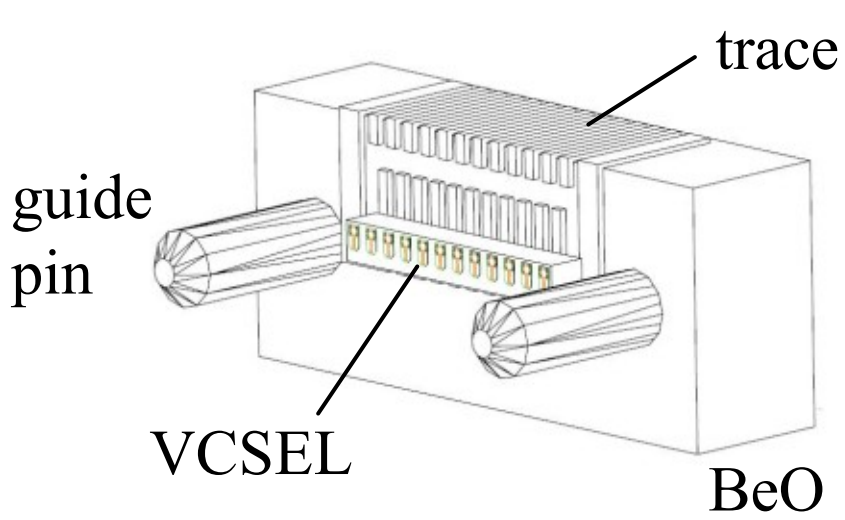

Figure 6: An optical package based on BeO.

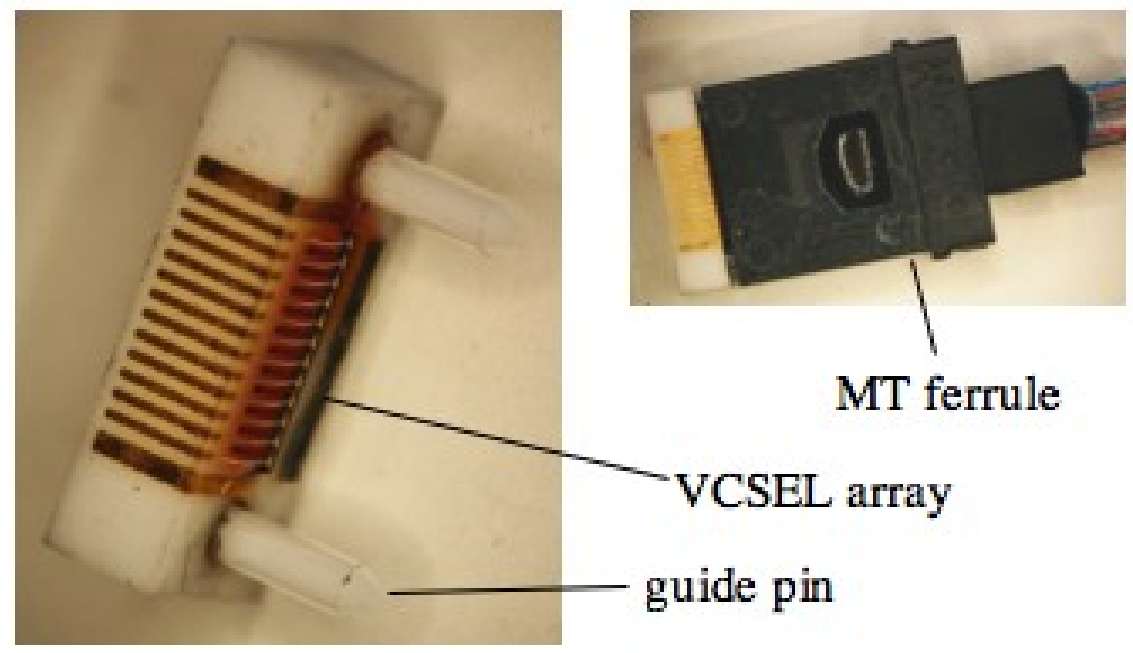

Figure 7: A fabricated opto-pack based on BeO.

The precise alignment of a VCSEL array to a MT ferrule is critical to achieve good optical power coupling; the alignment of a PIN array is much less critical because of the relatively large light sensitive area. Since the fibre ribbon is precisely placed with respect to the holes of the two guide pins in a MT ferrule, we align the VCSEL with respect to the guide pins. As a first step in the fabrication process, the guide pins are attached to the BeO base using epoxy with the precise relative location fixed by a MT ferrule. A VCSEL or PIN array is then aligned with respect to the guide pins under a microscope. We achieve good coupled optical power for the VCSEL arrays from various vendors. This demonstrates the principle of a compact opto-pack fabricated with BeO for heat management.

## VI. Summary

We have studied the bandwidth of the electrical and optical transmission lines of the current optical link of the ATLAS pixel detector. The results indicate that the micro twisted-pair cables can transmit signals up to 1 Gb/s. The fusion spliced fibre ribbon can transmit over 2 Gb/s. The GaAs VCSEL arrays from three vendors have been found to have the radiation hardness suitable for the SLHC operation. The silicon PIN arrays by Truelight are also found to be radiation-hard. The current ATLAS pixel optical link architecture can therefore be used at the SLHC as a possible upgrade scenario. We have also demonstrated the feasibility of fabricating a novel opto-pack for housing VCSEL and PIN arrays with BeO as the substrate.

## Acknowledgements

The authors are indebted to M. Glaser for the help at the T7 irradiation facility at CERN. This work was supported in part by the U.S. Department of Energy under contract No. DE-FG-02-91ER-40690.

## References

[1] K.E. Arms et al., “ATLAS Pixel Opto-Electronics,” Nucl. Instr. Meth. A 554, 458 (2005).
[2] ATLAS Pixel Detector Technical Design Report, CERN/LHCC/98-13.
[3] See, for example, J.H.R. Schrader et al, IEEE-J. Solid-State Circuits 41, 990 (2006).
[4] MWS Wire Industries Twistite.
[5] I. Gregor, “Optical Links for the ATLAS Pixel Detector”, Ph.D. Thesis, University of Wuppertal, (2001).
[6] A. Van Ginneken, “Nonionzing Energy Deposition in Silicon for Radiation Damage Studies,” FERMILAB-FN-0522, Oct 1989, 8pp.
[7] A. Chilingarov, J.S. Meyer, T. Sloan, Nucl. Instrum. Meth. A 395, 35 (1997).
[8] The fluences include a 50% safety margin.
[9] http://irradiation.web.cern.ch/irradiation/irrad1.htm
[10] M.-L. Chu et al., Nucl. Inst. Meth. A 530, 293 (2004).